\documentclass[prl,twocolumn]{revtex4}
\usepackage[dvips]{graphics,color}
\usepackage{amsmath}
\usepackage{amsfonts}

\usepackage{amssymb}
\usepackage{amstext}
\usepackage[sort&compress]{natbib}
\newcommand{\ignore}[1]{}

\newcommand{\ket}[1]{| #1 \rangle}
\newcommand{\bra}[1]{\langle #1 |}
\begin{document}

%%%%%%%%%%%%%%%%%%%%%%%%%%%%%%%%%%%%%%%%%%%%%%%%%%%%%%%%%%%%%%%%%%%%%%%%%%%%%%%
\title{Generalised state spaces and non-locality in fault tolerant quantum computing schemes.}
%%%%%%%%%%%%%%%%%%%%%%%%%%%%%%%%%%%%%%%%%%%%%%%%%%%%%%%%%%%%%%%%%%%%%%%%%%%%%%%

\author{N. Ratanje and S. Virmani}

\affiliation{Department of Physics SUPA, University of Strathclyde, John Anderson Building,
107 Rottenrow, Glasgow, G4 0NG, United Kingdom}

\date{\today}

\begin{abstract}
We develop connections between generalised notions of entanglement and quantum computational devices where the measurements available are restricted, either because they are noisy and/or because by design they are only along Pauli directions. By considering restricted measurements one can (by considering the dual positive operators) construct single particle state spaces that are different
to the usual quantum state space. This leads to a modified notion of entanglement that can be very different to the quantum version (for example, Bell states can become separable). We use this approach to develop alternative methods of classical simulation that have strong connections to the study of non-local correlations: we construct noisy quantum computers that admit operations outside the Clifford set and can generate some forms of multiparty quantum entanglement, but are otherwise classical in that they can be efficiently simulated classically and cannot generate non-local statistics. Although the approach provides new regimes of noisy quantum evolution that can be efficiently simulated classically, it does not appear to lead to significant reductions of existing upper bounds to fault tolerance thresholds for common noise models.
\end{abstract}

\maketitle

A scalable quantum device is useful for non-classical computation if it cannot be simulated
efficiently classically, as any such device can always be used compute its own observable behaviour.
It is hence important to understand when a quantum device
can or cannot be efficiently simulated classically. In one direction this problem is hard: while it is conjectured to be the case that ideal quantum computers are not classically tractable,
proving this rigorously would resolve longstanding open problems in theoretical computer
science.

The opposite direction is usually more amenable to attack. For example, it is possible to construct various families of quantum system that can be efficiently simulated classically \cite{class0}, including ones that contain significant amounts of multiparty entanglement \cite{GK,class}. Of particular interest are the noise levels required before a proposed quantum computer becomes so noisy that it can be efficiently simulated classically \cite{ABnoisy}. These noise levels are important because they represent the noise levels at which the power to do non-classical computation is lost, and hence are strongly related to the topic of {\it fault tolerant quantum computation}. Studies in that area have shown that for a variety of noise models there is a so-called {\it quantum fault tolerant threshold} noise level, below which it is possible for a noisy quantum computer to simulate an ideal quantum computer efficiently. However, in almost all cases
the values of this threshold are unknown. {\it Lower bounds} to the thresholds have been obtained using constructive methods for quantum error correction that are robust to imperfections (see e.g. \cite{AB,Knill}). {\it Upper bounds} have been obtained
in two ways - one approach is to argue that once the noise is too high the output cannot contain sufficient information about the input \cite{Kempe,Kay,Razborov}, another approach is to argue that
above a certain noise level the system can be efficiently simulated classically \cite{ABnoisy,HN,VHP05,PV,Buhrman,Ben magic,Howard,earl}. It is this
latter `classical' approach that we will follow here.

Within the classical approach there are two methods for simulating noisy quantum computers that
will be relevant to us. The first is a method of Harrow \& Nielsen (HN) \cite{HN}, which argues that if the noise level is too high then the entangling gates in the device (for instance the CNOT gates) will
become non-entangling, and that once this happens the probability distribution of measurement
outcomes can be efficiently sampled classically. The second method uses the Gottesman-Knill (GK)
theorem \cite{GK} to show that if the noise levels are high enough, all the elementary operations in the devices will enter the so-called
{\it stabilizer} or {\it Clifford} set, and hence can be efficiently simulated using the stabilizer formalism.
The GK theorem is especially relevant for fault tolerant quantum
computation because the most well understood fault tolerance proposals are built using stabilizer operations
with an additional non-stabilizer resource.

The aim of the present work is to use a generalised notion of entanglement \cite{Barnum}
(in a sense described in the next section) to apply a modified version of the Harrow \& Nielsen method to devices where there are restrictions on the measurements available. Such restrictions arise naturally in fault tolerant quantum computation for two reasons. Firstly, most fault tolerance schemes only use measurements of certain observables (in particular Pauli measurements or their generalisations in higher dimensions). Secondly, any measurements that are available are always prone to noise. It turns out that such restrictions can allow one to consider non-quantum single particle state spaces, and thereby a new notion of entanglement.

We will show that this approach forms a bridge between the foundations of quantum theory and the study of noisy quantum computation. For instance, the approach shows that for quantum computers using physical measurements from the Pauli operators only, non-classical computation is only possible if the dynamics is either non-positive w.r.t. the non-quantum state space, or can generate non-local correlations. It also motivates the construction of generalised probabilistic theories \cite{Boxworld} that are inspired by fault tolerant quantum computation schemes (although in our primary case the generalised theory turns out to be simply classical), and provides regimes of classically tractable noisy quantum evolution that appear to fall outside the scope of existing methods. It also falls under the programme initiated in \cite{Barnum} to investigate so-called generalised notions of entanglement, and motivates some interesting questions regarding the contrast between different forms of generalised multiparty entanglement.

Any readers not interested in technical details may ignore the appendices and skim the detailed calculations in the latter half of the paper. In the first few sections we begin with some general motivation and properties of the restriction to Pauli measurements.

\section{Generalised notions of entanglement, and application to the Stabilizer formalism.}

For our purposes a classical simulation of a quantum device is a classical algorithm that samples the probability distribution of measurement outcomes that occur in the device (we
do not require the ability to calculate these probabilities directly). For a general quantum system this
sampling is believed to be impossible to achieve efficiently,
due to the conjectured superiority of the quantum computer.
However, if the quantum gates in the system are so
noisy that they are incapable of entangling separable input states,
then Harrow \& Nielsen (HN) \cite{HN} showed that it can be efficiently simulated
classically. It is helpful to illustrate their method in order to motivate the rest
of this note. Consider a noisy two-particle gate that
has this property (i.e. it is incapable of entangling separable inputs) and suppose
that it acts on the first two input qubits in the device. As the
 gate is by hypothesis separable, it will give an output state $\rho$ that is also separable \cite{Werner89}:
\begin{equation}
\rho = \sum_i p_i a_i \otimes b_i, \label{sep}
\end{equation}
where $a_i,b_i$ are states on the first and second qubits respectively, and $p_i$ is a probability distribution.
In such cases the algorithm of \cite{HN} does not store the whole state, a task which could require exponentially
 large memory, but instead classically samples the probability distribution $\{p_i\}$, and then only stores $a_j,b_j$ corresponding to the result $j$ of this sampling. This
  process is then repeated for every gate in the quantum circuit. When the last gate in the circuit
  has been applied, the computer memory will have in storage $N$ single qubit states (where $N$ is the number of qubits), requiring modest memory. To complete the simulation we sample what happens when single qubit measurements are performed on these states. It is intuitively clear that this process, if performed to sufficiently high accuracy, will lead to a final outcome that is sampled from the same distribution as would be obtained if the quantum device were actually built. A complete analysis of the algorithm requires an analysis of
  the errors induced by finite precision storage and sampling methods on classical computers. However,
  these details will not concern us, as the analysis presented in \cite{HN}
applies with only minor modification to the alternative notion of entanglement that we consider below. It is also
useful to note that the HN method also applies to adaptive schemes, where measurement outcomes are use to determine
future choices of gate. This property also carries over to the generalised notions of entanglement that we now consider.

The ordinary notion of quantum separability relies on the conventional notion of a single particle quantum state space. In a manner similar to \cite{Barnum}, we will consider a redefinition of the single particle state space away from the `quantum' version,
thereby giving a different notion of `separability' under which one can apply the algorithm
of \cite{HN}. As a starting point for our generalised state spaces we will use the following characterisation of the quantum set $Q_d$ of dimension $d$ single particle states. We may define $Q_d$ as the set of complex $d \times d$ matrices that give valid probability distributions for all POVMs using the Born rule:
\begin{equation}
Q_d = \{\, \rho \, | \, \, {\rm{tr}}\{M \rho\}\geq 0 \,\, , \forall I \geq M \geq 0 \} \label{states}
\end{equation}
This definition can be generalised: we may define a single particle `state space' as the set of matrices that give valid probability distributions for the set of measurements that we have available. If the set of measurements is restricted, then the set of `states' defined in this way will be larger than quantum - it will contain matrices that are not admissible as quantum states because they needn't be positive for measurements outside the restricted set. If we are considering quantum devices with such restricted measurements, then we may apply the HN algorithm to efficiently simulate gates that are separable w.r.t. these modified state spaces - the modified state spaces still allow us to sample the probability distribution obtained from the restricted measurements available. As the gates that are separable will be different to the quantum-separable ones, this leads to different regimes of classically efficient simulation.

There are various natural ways to select a set of ``allowed" measurements. One possibility is to consider a single POVM (which may or may not be noisy), and then construct a set of measurements by acting upon it with a symmetry group of single particle unitaries. A prototypical example is inspired by stabilizer architectures for fault tolerant quantum computation: in these proposals the measurements
are restricted to the Pauli axes, which can be generated from a single Pauli measurement using the `Clifford group' \cite{GK}. It is this restriction that we consider in the next section.

\section{Bloch Cubes}

\begin{figure}[t]
        \resizebox{6cm}{!}{\includegraphics{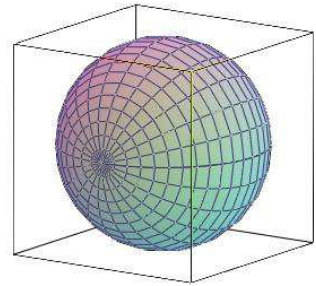}}
        \caption{The Bloch sphere represents the set of positive operators when all POVMs are included. If the Born
        rule is used with only Pauli measurements, however, then a cube of operators becomes positive. The eight corners
        of the cube are given by Bloch vectors $(s,t,u)$ with $s,t,u \in \pm 1$, and attain any set of possible outcomes
        for Pauli measurements. Hence we describe Pauli measurements as `operator compatible'. Any pure quantum state that is
        not a Pauli eigenstate lies on the surface of the sphere but strictly inside the cube.}
\label{cube}
\end{figure}

We begin by analysing the restriction to Pauli $X,Y,Z$  measurements.
It is easy to show that in this case the set of valid states is a `cube', the corners of which are
given by eight Bloch vectors
\begin{equation}
\{(1,1,1),(1,1,-1),(1,-1,1),..,(-1,-1,-1)\}.
\end{equation}
To see this, note that because the Pauli matrices form a basis, any
2 dimensional matrix can be written as
\begin{equation}
{1 \over 2}(aI+bX+cY+dZ)
\end{equation}
for some expansion coefficient $a,b,c,d$.
Under the Born rule, probability of getting +1 on measuring $Z$ is $(a+d)/2$, and the
 probability of getting the -1 outcome is $(a-d)/2$.
Hence for $Z$ measurements alone any matrix with $a=1$ and $d \in [-1,+1]$ gives a valid probability
distribution. Similar conclusions hold for $X,Y$ measurements, and so we find that
any matrix in the `Bloch cube' with $a=1$ and $b,c,d \in [-1,+1]$ is admissible (see fig. (\ref{cube}), and also
reference \cite{galvao} for some related aspects of this geometry for hidden variable models).

Now that we have established these cubes of `single particle states' for $X,Y,Z$ measurements,
let us consider the implications of using them to define a modified notion of entanglement. We
define a separable two particle state as one that is a convex combination of products
of valid ``single particle states". i.e. a separable two particle state is now one that can be written
as
\begin{equation}
\rho = \sum_i p_i a_i \otimes b_i.
\end{equation}
where $a_i,b_i$ are single qubit operators represented by vectors from the Bloch cube, not
just positive states from the Bloch sphere.

If we have a quantum device that can only perform Pauli
$X,Y,Z$ measurements, and if the gates in the device take separable Bloch cube input states to
cube-separable output states, then we can apply the HN algorithm without modification
to efficiently simulate the system classically. This is particularly relevant for the `magic state' architectures \cite{BK,KLZ}. These devices
have the following attributes:
\begin{itemize}
\item The ability to measure in the Pauli bases.
\item The ability to prepare single qubit states that are not mixtures of Pauli eigenstates.
\item The ability to do single qubit unitaries from the Clifford group - these unitaries are all symmetries
of the Bloch-cube.
\item The ability to do a single two qubit operation, typically a CNOT or CSIGN gate, or something else sufficient
for generating the entire Clifford group on n-qubits. In these notes we shall only consider the CSIGN gate,
which is represented by the unitary matrix $\ket{00}\bra{00}+\ket{01}\bra{01}+\ket{10}\bra{10}-\ket{11}\bra{11}$.
The results that we will obtain can be applied to other entangling Clifford group gates by modifying the local $X,Y,Z$ reference
frame appropriately.
\end{itemize}
The first three of these attributes fit nicely into the picture of `cube states' - the Pauli directions are the
allowed measurements, single qubit states from the Bloch sphere are contained entirely within the Bloch-cube,
and the single qubit Clifford unitaries are symmetries of the Bloch-cube.

The only operations that fail to preserve cube-separability are the last ones - the two qubit Clifford
operations. That this is true is a special case of a result in \cite{gross} in the context of box-world generalised theories,
but for convenience  in Appendix II we provide an alternative proof that is tailored to our problem.

We hence must answer the following question: {\it how much noise does it take to make a CSIGN gate cube-separability
preserving}? If we have such noise levels then we know that we can simulate the system efficiently classically,
and hence these noise levels should supply upper bounds to the fault tolerance thresholds.

Because the regime
covered by cube-separability is different from both the GK regime (we admit all single qubit states), and the
the original HN approach (there are entangled states that are cube-separable), there is a chance
that it can simulate some systems efficiently classically that are not covered by naive applications of the other two approaches.
We provide evidence for the existence of such regimes below, although it is possible that more sophisticated versions of these
existing algorithms (perhaps allowing entangled `clusters', as in \cite{ABnoisy}) could cover any regime that we propose here.

In considering non-quantum states, we are exploring a type of generalised probabilistic theory \cite{Boxworld} that can lie
outside conventional quantum mechanics. However, we are still using a large amount of quantum structure:
while the cubes of states that we consider are effectively (3-measurement, 2-outcome) `boxes' from box-world theories \cite{Boxworld}, we take our dynamics directly from noisy quantum mechanics with a {\it given Hilbert space}. In some cases the generalised theory that we are trying to construct is nevertheless effectively classical --- most of the calculations in the paper in fact involve identifying noisy quantum gates that can be considered valid operations in a classical theory. However, it is because we are considering noisy quantum gates that we choose to retain the description of the cubes in terms of Pauli operators, rather than using tables of probabilities that are more natural in generalised theories.

\section{Contrast between cube-entanglement and quantum entanglement, connections to nonlocality}

The main technical problem underlying these investigations is to decide when a noisy quantum operation is cube-separable. This section presents a number of observations that are useful for understanding the context of this problem, and its relationship to ordinary quantum entanglement and non-locality.

\smallskip

{\bf Definition 1: Operator compatibility.} We define a set of quantum
measurements to be {\bf operator compatible} if one can find, for all combinations of measurement outcomes, an operator (usually a non-positive operator) that achieves those outcomes deterministically under the Born rule. $\blacksquare$

\smallskip

For instance, $X,Y,Z$ measurements are operator compatible: suppose that you wish to get values
$e,f,g$ for measurement of $X,Y,Z$ respectively, where $e,f,g \in \pm 1$. Then the operator
$1/2(I+eX+fY+gZ)$ achieves this under the Born rule. These operators are just the corners of the Bloch cube. The same argument extends to any set of measurements described by traceless orthogonal observables (in Hilbert-Schmidt inner product), as one can write out the desired operator as a suitable linear sum of the observables.

While this notion of `compatibility' for sets of quantum measurements does not occur in general, it is not difficult to generically characterise the set of operator compatible measurements:

\smallskip

{\bf Lemma 2: Characterisation of operator compatible measurements.} On quantum systems
of dimension $d$ a set of $N$ distinct POVMs can only be operator compatible if the total number
of outcomes summed over all the POVMs is $\leq d^2 + N - 1$. Generically chosen measurements satisfying this constraint
are operator compatible.

\smallskip

{\it Proof:} Consider $N$ POVMs $P_1,P_2..,P_N$ of a $d$ dimensional quantum system, and suppose
that these POVMs have $m_1,m_2..,m_N$ outcomes respectively. A trace-one hermitian matrix $A$ of dimension $d \times d$ has $d^2-1$ free parameters. Under the Born
rule only one of the POVM elements in each POVM must have probability one, the rest must have probability zero. However, one of these $m_i$ constraints is guaranteed by normalisation. Hence each measurement $P_i$ leads to $m_i-1$ independent linear constraints on $A$. Hence we will generically require that:
\begin{eqnarray}
&& \sum_{i=1,..,N} m_i-1 \leq d^2 - 1 \nonumber \\
\Rightarrow && \sum_{i=1,..,N} m_i \leq d^2 + N - 1 \,\,\,\,\,\, \blacksquare \nonumber
\end{eqnarray}

\smallskip

In particular, if there are $d$ measurement outcomes for each measurement, and if $N > 1$, then this condition reduces to
(for $N=1$ the condition is trivial anyway):
\begin{eqnarray}
&& 0 \leq d^2 - N.d + (N-1) \,\,\,\,\,\, \nonumber \\
\Rightarrow && 0 \leq (d-(N-1))(d-1) \,\,\,\,\,\, \nonumber \\
\Rightarrow && N \leq d+1 \,\,\,\,\,\, \nonumber
\end{eqnarray}

\smallskip

The property of operator compatibility has important consequences if the measurements being considered are
{\it tomographically complete}, i.e. if the values of the probabilities of each measurement outcome are sufficient to fix the state.
In these cases a matrix is separable w.r.t. to the single party state space (as constructed via equation (\ref{states})) iff it
satisfies a local hidden variable model:

\smallskip

{\bf Lemma 3: A multiparticle state is separable w.r.t. the state spaces corresponding to tomographically complete operator compatible measurements iff there is a local hidden variable model for those measurements made upon the state.}

\smallskip

{\it Proof:} We consider only two particles for simplicity, the argument straightforwardly to multiparticle systems. As the measurements are tomographically complete, this means that a state $A$ fixes uniquely and is uniquely fixed by the traces of the matrix with the measurement operators. Hence the state is exactly characterised by the probability
 distribution of measurement outcomes.

If the probability distribution satisfies a local hidden variable model \cite{LHV}, then
by definition it can be written as
\begin{eqnarray}
{\rm{tr}}\{A ((P_a)_i \otimes (P_b)_j ) \} = \sum_{\lambda} p(\lambda) p(i|a,\lambda)q(j|b,\lambda)
\end{eqnarray}
where $(P_a)_i$ and $(P_b)_j$ are the $i$th and $j$th elements of POVMs $P_a,P_b$ respectively, $p(\lambda)$ is a probability distribution over a hidden variable, and $p(i|a,\lambda)$ and $q(j|b,\lambda)$ are
probability distributions for local outcomes. These local probability distributions can in turn be decomposed
as mixtures of local {\it deterministic} distributions, where a single deterministic outcome is specified
for each measurement choice. The deterministic distributions in turn are matched by the extreme points of
the state space (as defined via eq. (\ref{states})), and so a matrix $A$ is separable w.r.t. the state space
defined by the measurements iff it satisfies a local hidden variable model for those measurements.
$\blacksquare$

\smallskip

In particular for the case of qubit Pauli measurements a quantum state is cube-separable iff it has a local hidden variable model for Pauli measurements. If a set of allowed measurements is not `operator compatible', then this opens up the possibility
that there is a distinction between non-locality and non-separability - just as has been debated in the setting of conventional quantum theory \cite{Werner89}.

The connection between hidden variable models and non-classical computation is still an unresolved question (see e.g. \cite{Rauss} for other investigations in this direction). The property of `operator compatibility' sheds partial light on this question for machines restricted to Pauli measurements. As we can classically efficiently simulate any dynamics that preserve separability of the `cubes' corresponding to the operator compatible measurements, we have that:

\smallskip

{\bf Corollary 4: Quantum devices using only tomographically complete operator compatible measurements cannot perform non-classical computation if the quantum dynamics considered as dynamics of the modified state spaces cannot generate negative or non-local measurement probabilities (as computed via the Born rule) when acting upon states from the modified state space.}

\smallskip

Note that in this corollary we need to exclude the possibility of having negative probabilities because even single particle unitaries (not from the Clifford group) can rotate cube states to matrices that are not within the cubes.

\bigskip

\smallskip

{\bf Lemma 5: Bell states of 2-qubits are cube separable, hence there are quantum operations that are cube-separable, despite being entangling.}

\smallskip

{\it Proof:} It is well known that Bell states have a LHV model for $X,Y,Z$ measurements (one way of proving this is presented later). Hence any quantum operation
that creates Bell-diagonal two qubit states (e.g. by throwing the input
away and creating a bell state in its place) will be separable in the
cube setting, even though may be entangling in the usual quantum setting. $\blacksquare$

\smallskip

While Lemma 5 establishes a difference between cube-separable operations and entangling operations, operations that create Bell diagonal states are {\it entanglement breaking} (they break entanglement between the
two-qubits and the rest of the universe \cite{EBdef}) and so they cannot be used to generate multiparty entanglement
by themselves (and indeed they can be classically efficiently simulated \cite{VHP05}). In Lemma 8 we will present
examples of quantum operations that are cube-separable despite being capable of generating
some form of multi-particle entanglement. However, first let us observe the following:

\smallskip

{\bf Lemma 6: Around almost any quantum product pure state of many qubits, there is a region of cube-separable states.}

\smallskip

{\it Proof:} Consider a pure quantum state $\sigma$ that is not an eigenstate of the Pauli $X,Y,Z$ operators. Such a state can be represented
as a mixture of cube-pure state with strictly non-zero weights for all cube-pure states. Consider a small
Hermitian perturbation $\Delta$ to products of such pure quantum states, to give an operator
$\sigma + \Delta$. As cube-pure states form an overcomplete
basis for the space, $\Delta$ can be expanded as a real combination of products of cube-pure states, possibly with negative weights.
However, if the norm of $\Delta$ is small enough, these negative weights will be small enough not to
take the product pure state $\sigma$ out of the cube separable states (each non-Pauli pure state lies on the Bloch sphere but is strictly inside the cube - see Fig(\ref{cube})). $\blacksquare$

\smallskip

A quantum state is said to be genuinely $N$-party entangled if it cannot be written as a probabilistic mixture of
states that are products of entangled states of less than $N$ parties. In the case of three parties, for example, it has been shown \cite{Wstate} that
there are two classes of 3-qubit entanglement - the `W' and GHZ class entangled states, and that examples of such states can
be found that are arbitrarily close to product states. Hence we have the following observation:

\smallskip

{\bf Corollary 7: There are genuinely multi-particle quantum-entangled states that are cube-separable.}

\smallskip

{\it Proof:} Both $W$ and $GHZ$ class 3-qubit states can be found that are arbitrarily close
to product states. Hence the previous lemma shows that there are $W$ and $GHZ$ class entangled
states that are cube-separable. $\blacksquare$

\smallskip

This argument suggests that it is quite likely that a variety of genuinely multipartite entangled states can be found that are cube separable, and hence have a LHV model for X,Y,Z measurements. However, as the
number of qubits increases the size of the cube-separable ball will decrease.

However, as we show in the next lemma, it is possible to construct noisy quantum gates that are not quantum-separability preserving
(i.e. they can entangle product input quantum states), are not entanglement breaking, and that are not Clifford operations. The first two
of these properties mean that these operations can act on systems of product input particles to generate entanglement of increasing numbers of particles. The final property means that these entangled states can be chosen to be non-Clifford. However, it is not yet clear to us whether the entanglement generated by these cube-separable but not quantum-separable operations can become arbitrarily long range, and this
is something that warrants further investigation. It is important to note however that quantum Bell states are cube-separable, so they can
in principle be included as a free resource to help generate multiparticle entanglement.

\smallskip

{\bf Lemma 8: There are noisy quantum gates that are cube separability preserving but not (a) quantum separability preserving, (b) entanglement breaking, or (c) Clifford operations.}

\smallskip

{\it Proof:} There is a semi-systematic way of generating a family of quantum operations with this property.
Consider the magic `T-state' \cite{BK} - the qubit with Bloch vector $\sqrt{1/3}(1,1,1)$. Let this
state be denoted $\ket{T}$ and the orthogonal qubit state be denoted $\ket{\bar{T}}$. Then we may construct
a Choi-Jamiolkowski (CJ) \cite{CJ} state for a 2-qubit quantum operation acting upon qubits $A,B$ as follows. Let $A1,A2$ be the two qubits
of the CJ state corresponding to qubit $A$, and $B1,B2$ be the two qubits of the CJ state corresponding
to qubit $B$. Consider CJ state made by mixing with probabilities $1/2 + \epsilon$ and $1/2 - \epsilon$ respectively the following two
normalised states:
\begin{eqnarray}
(\alpha \ket{T_{A1}T_{A2}T_{B2}} + \beta \ket{\bar{T}_{A1}\bar{T}_{A2}\bar{T}_{B2}}) \otimes I_{B1}/2 \nonumber\\
(\gamma \ket{\bar{T}_{A1}T_{A2}\bar{T}_{B2}} + \delta \ket{T_{A1}\bar{T}_{A2}T_{B2}}) \otimes I_{B1}/2
\label{counter}
\end{eqnarray}
where we abuse notation slightly to keep the notation uncluttered - the first terms are written in `coherent' form, whereas
the state on particle $B1$ is written in operator form - ideally we should have written both as operators.

However, it is not difficult to show that provided that
\begin{equation}
({1 \over 2}+\epsilon)|\alpha|^2+({1 \over 2}-\epsilon)|\delta|^2 = {1 \over 2} \label{constraint}
\end{equation}
that the reduced state
of particles $A1,B1$ is maximally mixed. Hence the mixture is a valid CJ state for a trace preserving quantum operation
\cite{CJ}. But if both $|\alpha|,|\gamma|$ are close enough to 1,
the output state of this operation is close to a mixture of $T_{A2}\bar{T}_{B2}$ and $T_{A2}T_{B2}$, regardless
of the input state. By linearity this extends to input cube-states that are not quantum -- so for all input cube states, the output states will be close to a mixture of $T_{A2}\bar{T}_{B2}$ and $T_{A2}T_{B2}$. This means that provided that we pick $|\alpha|,|\gamma|$ close enough to 1
while maintaining condition (\ref{constraint}) (which we can do provided that $\epsilon$ is close enough to 0) the operation is
cube-separable, because as explained earlier, states close enough to non-Pauli pure product states (and hence states close to mixtures of $T_{A2}\bar{T}_{B2}$ and $T_{A2}T_{B2}$) are cube-separable. The output qubit $A2$ is also always close to $T_{A2}$, and hence the quantum operation represented
by equation (\ref{counter}) is not a Clifford operation, as together with the Clifford resources required to do teleportation it can generate $T_{A2}$.

We would like to argue that the operations in equation (\ref{counter}) can be chosen such that they are neither entanglement breaking nor separability preserving. They definitely aren't entanglement breaking or separable because creating the CJ state involves generating
 entanglement -- measuring $T,\bar{T}$ parity on A2B2 of the CJ state distills a GHZ-like state, similarly measuring
 parity on A1A2 does the same. Hence the operation is neither separable nor entanglement breaking
 as it is entangled across both the $1:2$ split and the $A:B$ split.

However, to ensure that the gates that we define do not fall under the regime covered by the Harrow \& Nielsen
algorithm \cite{HN}, we must also show that they are not {\it separability-preserving}, i.e. we must show that there are input product
states that are entangled by the gate, even without the use of extra ancillas that would be required to make
the CJ state. It can be shown by explicit calculation (using the PPT criterion \cite{ppt}) that any product quantum input states with
the $A$ particle in $1/\sqrt{2}(\ket{T}+\ket{\bar{T}})_A$ are taken to non-PPT outputs provided that $\epsilon > 0$. $\blacksquare$

\section{Dynamical quantum gates and notation for representing two-2level operators}

Most of the remaining sections of the paper will be concerned with determining the noise levels required to
turn a CSIGN gate into a separable operation with respect to the modified notions of entanglement that we consider.
In this section we lay out some of our notation. The same notation will be used both for the analysis of cube-separability,
as well as the analysis of the rescaled Bloch-sphere that appears later.

For our purposes the action of the CSIGN gates is most easily represented by its action on Pauli operators.
It is hence convenient to represent the state of
two cubes in the basis of products of Pauli operators. Let $\sigma_0 = I, \sigma_1 = X, \sigma_2 = Y$, and
$\sigma_3 = Z$. Any two-cube operator can be represented as:
\begin{equation}
A = \sum_{ij} A_{ij} \sigma_i \otimes \sigma_j
\end{equation}
It is convenient to represent the set of coefficients $A_{ij}$ as a 4 x 4 matrix
where the matrix elements refer to the coefficient of the Pauli operator expansion as follows:
\begin{eqnarray}\left(\begin{array}{cccc}
     II & IX & IY & IZ \\
     XI & XX & XY & XZ \\
     YI & YX & YY & YZ \\
     ZI & ZX & ZY & ZZ
      \end{array}\right) \nonumber
\end{eqnarray}
e.g. the row 1, column 2 matrix element is the coefficient of the $I \otimes X$ operator
in the expansion. For example, an explicit calculation shows that the quantum Bell state $1/\sqrt{2}(\ket{00}+\ket{11})$ has the expansion:
\begin{eqnarray} {1 \over 4} \left(\begin{array}{cccc}
     1 & 0 & 0 & 0 \\
     0 & 1 & 0 & 0 \\
     0 & 0 & -1 & 0 \\
     0 & 0 & 0 & 1
      \end{array}\right) \nonumber
\end{eqnarray}
whereas a product state of two cube states is given by:
\begin{eqnarray} {1 \over 4} \left(\begin{array}{cccc}
     1 &  x_2 & y_2 & z_2 \\
     x_1 & . & . & . \\
     y_1 & . & . & . \\
     z_1 & . & . & .
      \end{array}\right) \nonumber
\end{eqnarray}
where the ``.'' elements are given by the product of the first element
of the row, and the first element of the column. - e.g. element
(2,3) in the matrix is given by $x_1 y_2$. The ``pure product'' states
in the cube picture are ones where all the $x_1,x_2,y_1,y_2,..$ are either
+1 or -1. In many cases we will ignore normalisation factors such as $1/4$.

This representation provides one way of seeing that the Bell state $1/\sqrt{2}(\ket{00}+\ket{11})$ is separable in the
cube picture - it is given by a uniform average of all cube pure
product states that satisfy the constraint $x_1=x_2,y_1=-y_2,z_1=z_2$
(this then extends to all the Bell states, because they are related to each other by Pauli unitaries on
one side).

Most of the calculations involving cubes in this paper will involve first ensuring that the output
of a noisy CISGN leads to a valid general theory on the cube-states. The arguments proceed along the following lines.
Suppose that you have a product of two cube-states
input into a two-cube linear operation such as a CSIGN. If we want the theory to be self consistent, we have
to make sure that the output state from this process gives a valid probability distribution
for the set of allowed $X,Y,Z$ measurements. A CSIGN acts in the following way upon a product input state:
\begin{eqnarray}{1 \over 4}\left(\begin{array}{cccc}
     1 & A & B & C \\
     x & xA & xB & xC \\
     y & yA & yB & yC \\
     z & zA & zB & zC
      \end{array}\right) \rightarrow {1 \over 4}\left(\begin{array}{cccc}
     1 & zA & zB & C \\
     xC & yB & -yA & x \\
     yC & -xB & xA & y \\
     z & A & B & zC
      \end{array}\right) \nonumber
\end{eqnarray}
On such an output state, what is the probability of measuring Pauli product $P \otimes Q$,
and getting say the outcome +1 for $P$ and -1 for $Q$? Well we use the usual Born rule:
\begin{equation}
{1 \over 4}{\rm{tr}}\left\{ ( (I + P) \otimes (I-Q) )     \rho           \right\}
\end{equation}
which is just given by an appropriate sum of four elements from the matrix of coefficients
representing the output state. So if for example $P=X$ and $Q=X$, then we just take a linear combination of
the top left corner matrix elements:
\begin{equation}
{1 \over 4} ( 1 + xC - zA - yB)
\label{cond}
\end{equation}
If the theory is self consistent, then we need all expressions such as this to
be valid probabilities, and so they cannot be negative. In Appendix II we show that without adding noise, the CSIGN gate
acting on any state spaces larger than the Bloch sphere leads to negative values for such expressions. As cube-separability
implies positive probabilities for Pauli measurements, this means that the noiseless CSIGN cannot be a cube-separable
operation.

\section{Noise models and corresponding bounds for cube-separability.}

The arguments of Appendix II show that a CSIGN gate cannot be cube-separable, as it does not even lead
to positive output states (for the allowed Pauli measurements) when acting upon input product cube states. However, as we shall see, it can often be made cube separable by the addition of noise. In this section we consider a few standard noise models on the CSIGN gate, and bound or compute the noise levels required to make it cube-separable. One might hope that because of example (\ref{counter}), there may be situations involving natural noise models for which the consideration of cube-separability may lead to tighter upper bounds on non-classical computation thresholds.

However, we shall see that in all the cases considered in this section the upper bounds derived are either the same or worse than those that can be derived by considering quantum separability, or have not been proven to be lower by us. In fact for magic state architectures (which were the primary motivation for considering the cubes anyway) better bounds can in many cases be obtained using the GK theorem \cite{foot}.

To show that a given (noisy) gate is cube-separable, one would have to show that all possible combinations
of input product pure cube states (the corners of the cubes) are taken to output states that are cube-separable.
However some of the noise models that we will consider have certain symmetry properties which mean that it is sufficient to consider only one input cube pure product state. In particular, suppose that we have a CSIGN gate $U$ acting on a pair of pure cube states $\rho$ and some CP map error $E$ acting with probability $\lambda$, such that under the total action of the noisy gate the output state is cube-separable:
\begin{equation}
(1-\lambda) U(\rho) + \lambda E(U(\rho)) \in {\rm{cube-separable}} \nonumber
\end{equation}
Now suppose that $E$ is such that it commutes with any local Pauli product unitary transformation $Q$, as well local products of
phase gates $S:=\ket{0}\bra{0}+i\ket{1}\bra{1}$, both acting by conjugation. Then it will be the case that:
\begin{eqnarray}
(1-\lambda) Q U(\rho) Q^{\dag} + \lambda  E( Q(U(\rho))Q^{\dag}) \in {\rm{cube-separable}} \nonumber \\
(1-\lambda) S U(\rho) S^{\dag} + \lambda  E( S(U(\rho))S^{\dag}) \in {\rm{cube-separable}} \nonumber
\end{eqnarray}
because both $Q$ and $S$ are Clifford unitaries and hence preserve the cubes. Now we can move $Q$ through the CSIGN $U$
to get another Pauli transformation $P$, and we can move $S$ through the CSIGN simply because it commutes:
\begin{eqnarray}
(1-\lambda)  U(P \rho P^{\dag}) + \lambda E(U( P \rho P^{\dag})) \in {\rm{cube-separable}} \nonumber \\
(1-\lambda)  U(S \rho S^{\dag}) + \lambda E(U( S \rho S^{\dag})) \in {\rm{cube-separable}} \nonumber
\end{eqnarray}
where $P$ is some other Pauli product. Now by choosing sequences of $P$ gates and $S$ appropriately,
we can take an cube extremal state to any other \footnote{ For instance starting with the cube Bloch vector
$(1,1,1)$ we can cycle through $(1,1,1) \rightarrow^X (1,-1,-1) \rightarrow^Y (-1,-1,1) \rightarrow^X (-1,1,-1)
\rightarrow^S (-1,-1,-1) \rightarrow^X (-1,1,1) \rightarrow^Y (1,1,-1) \rightarrow^X (1,-1,1)$. }.
This hence means that if $\lambda$ is strong enough
to make one particular input cube pure product state separable, it can make any cube state separable, provided
that $E$ satisfies the required commutation properties. If $E$ corresponds to local or joint dephasing or
depolarisation noise (as defined later), then it has these properties, and so for those noise
models we shall only consider what happens to one particular product input cube-state.

Despite this simplification the problem is still not easy in general, because deciding whether there is a LHV model for
 a given joint probability distribution is not straightforward \cite{web}. In the relevant case of 2 parties, 3 measurements, and 2 measurement outcomes, a complete but large list of Bell inequalities has been obtained \cite{PS}. However, we shall find
that for most of the noise models that we examine, the problem can be completely or almost completely solved
analytically -- in fact in most cases the key constraint is
not one of locality, but of having a valid probability distribution for the outcomes.
\begin{itemize}
\item {\bf Error-per-Gate noise on 2 particles undergoing CSIGN. The minimal error rate $\lambda$ required satisfies
$50\% \geq \lambda \geq 20\%$.}

In this noise model
 we apply a perfect CSIGN and then afterwards with probability $p$ apply a general
 {\it quantum} CP map of our choosing - we may be as adversarial as possible within the set
 of two-qubit CP maps. It can be seen that $p=$50\% noise is sufficient. To see that it is sufficient, consider dephasing one particle completely prior to performing the CSIGN gate -
this corresponds to 50\% $Z$ noise on one arm. This takes that particle to a mixture of $\ket{0}$ and $\ket{1}$,
which means that the second particle will be probabilistically rotated by $Z$. As $Z$s preserve the cube this means
that the overall operation will be cube-separable.

We may also show that at least 20\% noise is necessary
using the following argument. Consider the cube-extremal state given by Bloch vector $(1,1,1)$. We may write this
state as a linear combination of the density matrices $T$ and $\bar{T}$, which we define as the
density matrices corresponding to the magic states $\ket{T}$ and $\ket{\bar{T}}$ respectively. The expression is:
\begin{eqnarray}
{1 \over 2}(I + X + Y + Z) = \left({ \sqrt{3} + 1\over 2}\right)T - \left({ \sqrt{3} - 1\over 2} \right)\bar{T} \nonumber \\
= w T - (w-1) \bar{T} \nonumber
\end{eqnarray}
where we define $w:={ \sqrt{3} + 1\over 2}$. Similar expressions hold for the other cube extrema, albeit with different magic states. So the tensor product of two of cube extrema will be of the form:
\begin{eqnarray}
w^2 T_A \otimes T_B + (w-1)^2 \bar{T}_A \otimes \bar{T}_B \nonumber \\
- w (w-1) T_A \otimes \bar{T}_B - w (w-1) \bar{T}_A \otimes T_B \nonumber
\end{eqnarray}
for two choices of magic state $T_A,T_B$.
Hence any CP map acting upon the product of cube extrema inputs will give an
expression of the form:
\begin{eqnarray}
w^2 \rho_1 + (w-1)^2 \rho_2
- w (w-1) \rho_3 - w (w-1) \rho_4 \label{diff}
\end{eqnarray}
where $\rho_1,..,\rho_4$ are quantum states. Note that both $w$ and $w-1$ are positive,
and so the maximum `probability' for a measurement outcome on equation (\ref{diff})
will be no greater than:
\begin{eqnarray}
w^2 + (w-1)^2 = 2
\end{eqnarray}
Equation (\ref{cond}) with the choice $x=y=z=A=B=-C=1$ gives a negative probability of -1/2 for the pure CSIGN, so at the very least we need the noise to be sufficiently strong to make this positive. Now as the noise can at most balance this out by mixing in
a `probability' of $w^2 + (w-1)^2 = 2$, we require that:
\begin{eqnarray}
(1-\lambda)\left({-1 \over 2}\right) + \lambda (2) \geq 0 \nonumber \\
\Rightarrow \lambda \geq {1 \over 5} = 20\% \nonumber
\end{eqnarray}

We have not been able to obtain an exact value for the error-per-gate noise
required to make CSIGN cube-separable. However these values should be compared
to the noise required for the CSIGN to become quantum-separable, which is $50\%$ \cite{HN}.

\item {\bf Joint depolarising on output. A noise level of $\lambda \geq 2/3$ is necessary and sufficient.}

Joint depolarising noise on a 2-cube operator acts is parameterised by
a probability $\lambda$ and acts as $\rho \rightarrow (1-\lambda) \rho + \lambda {\rm{tr}}\{\rho\}I/4$ immediately
after the action of an ideal CSIGN. In terms of the matrix representations that we have been using above, this transformation
is represented as $A_{00} \rightarrow A_{00}$, otherwise $A_{ij \neq 00} \rightarrow (1-\lambda)A_{ij}$. It can be
 shown that $ \lambda = 2/3$ noise is both necessary and sufficient. Necessity follows from
equation (\ref{cond}): picking input cubes with $x=-C,z=A,y=B$ and then applying the depolarising noise,
 we find that requirement (\ref{cond}) is equivalent to $1-3(1-\lambda) \geq 0$, and hence $\lambda \geq 2/3$.
 That this level of noise leads to a system with a LHV is shown in the appendix by considering
instead an input state where $x=y=z=A=B=C=1$.

This is essentially the same as ordinary quantum case - as 2/3 total joint depolarisation is required to
take an EPR pair to a separable state. Hence the consideration of cube separability leads to no reduction in the upper bounds.

%However, it is interesting to consider what happens if pure Bell states are also available to the machine.
%In this case considerations of quantum-separability alone show that the CSIGN must be subject to 8/9 joint depolarisation
%\cite{HN} before becoming separable, as this is the noise required to remove the quantum-entanglement of
%the CJ state corresponding to the CSIGN. However, as Bell states are cube-separable we may admit them for free, and
%so the joint depolarisation rate of 2/3 still holds. Hence when given access to Bell pairs there is a reduction over
%the result that can be obtained by quantum separability. BUT WHAT ABOUT CLIFFORD?

\item {\bf Local depolarising noise. An error rate of $p \geq 2 - \sqrt{2} \sim 60\%$ is necessary and sufficient.}

Local depolarising of strength $\lambda$ acts as $\rho \rightarrow (1-\lambda) \rho + \lambda {\rm{tr}}\{\rho\}I/2$ independently on each particle. In terms of the Pauli coefficients $A_{ij}$ this corresponds
    to $A_{00} \rightarrow A_{00}$, $A_{0j} \rightarrow (1-p) A_{0j}$ for $j \neq 0$, $A_{i0} \rightarrow (1-p) A_{i0}$ for $i \neq 0$, and $A_{ij} \rightarrow (1-p)^2 A_{ij}$ otherwise. Applying condition (\ref{cond}) for CSIGN acting upon input pure
cube states with $x=-C,z=A,y=B$ followed by such noise gives
    \begin{equation} \nonumber
1 - 2(1-p) - (1-p)^2 \geq 0 \Rightarrow p \geq 2 - \sqrt{2} \sim 60\%
\end{equation}
That this level of noise leads to a system with a LHV is shown in the appendix by considering
instead an input state where $x=y=z=A=B=C=1$ - the relevant LHV is the last one presented
in the appendix.

This figure is worse than the local depolarising rate required to turn an arbitrary
entangled pure 2 qubit state into a separable state, which using the PPT criterion \cite{ppt} can
be easily shown to be $p = 1 - {1 \over \sqrt{3}}=42.2\%$.

\item {\bf Local dephasing noise. An error rate of $p \geq 1 - {1 \over \sqrt{2}} \sim 30\%$ is necessary and sufficient.}

Local dephasing of strength $p$ implements a $Z$ with probability $p$ on each particle
independently. In terms of the matrix representation that we have been using above, this is equivalent to
$A_{ij} \rightarrow A_{ij}$ whenever $i \in \{0,3\}$ and $j \in \{0,3\}$, $A_{ij} \rightarrow (1-2p)^2 A_{ij}$ whenever $i \in \{1,2\}$ and $j \in \{1,2\}$, and otherwise $A_{ij} \rightarrow (1-2p) A_{ij}$. Applying condition (\ref{cond}) for CSIGN acting upon input pure
cube states $x=-C,z=A,y=B$ followed by such dephasing noise hence leads to the condition:
\begin{equation} \nonumber
1 - 2(1-2p) - (1-2p)^2 \geq 0 \Rightarrow p \geq 1 - {1 \over \sqrt{2}} \sim 30\%
\end{equation}
That this level of noise leads to a system with a LHV is shown in the appendix by considering
instead an input state where $x=y=z=A=B=C=1$ - the relevant LHV is the penultimate one presented
in the appendix.

Unfortunately the value of $p \geq 1 - {1 \over \sqrt{2}} \sim 30\%$ is no tighter than can be obtained
by considering quantum-separability: it is precisely the same local dephasing rate as is required
make the CSIGN a quantum-separable operation (we omit the details to show this, but it can be shown by considering the CJ state of the noisy operation,
which turns out to be locally equivalent to a separable Bell diagonal state in an encoded basis $\ket{\bar{0}}=\ket{00},\ket{\bar{1}}=\ket{11}$).

\end{itemize}

\section{Noisy Measurements or Preparations: Rescaled Bloch sphere.}

An alternative scenario (and a very realistic one) in which measurements can be restricted is when they are noisy. In this section we shall find that admitting noise on the measurements allows the CSIGN gate to become separable (with respect to the redefined state space) with slightly less noise than is required for quantum separability. The differences are slight (Figs.(\ref{jointRasNoise}),(\ref{localRasNoise})), but do reveal some interesting behaviour. Firstly, one can consider adding noise to preparations instead of measurements, but we find that in contrast {\it more} noise is required. Secondly, in the case of local dephasing noise we find that the noisy CSIGN cannot be a valid dynamical operation (in a generalised theory sense) for the non-quantum state spaces unless the dephasing is maximal.

First let us establish the state spaces that we consider. Consider a given projective measurement $\{\ket{e}\bra{e},\ket{\bar{e}}\bra{\bar{e}}\}$, and suppose that immediately prior to it the particle being
measured is depolarised with probability $p$. The fact that the measurement is noisy
means that a longer Bloch vector can be admitted than in the noise free setting. A qubit matrix $\rho$
(perhaps from outside the Bloch sphere) will be admissible for a noisy version of the measurement if
\begin{eqnarray}
(1-p){\rm{tr}}\{\ket{e}\bra{e} \rho \} + {p \over 2} \geq 0 \nonumber \\
(1-p){\rm{tr}}\{\ket{\bar{e}}\bra{\bar{e}} \rho \} + {p \over 2} \geq 0
\end{eqnarray}
for all possible measurement directions. In the case of qubits this condition translates to the requirement that the Bloch vector
$v$ representing the state $\rho$ be valid in the noisy setting if
\begin{equation}
v = Rv'  \label{noisy1}
\end{equation}
defines a valid vector $v'$ in the noise-free setting, where $R$ is given by:
\begin{equation}
R = {1 \over 1 - p}
\end{equation}
This simply means that the new noisy valid set is equal to the noise-free valid
set enlarged by a factor $R > 1$. Actually the rescaling factor $R$ can be used to describe
noise in the preparation of qubits too: if we have $R < 1$ then this corresponds to preparing
qubits and then depolarising them individually at a rate $1-R$ (i.e. $\rho \rightarrow R \rho + (1-R)I/2$).
Hence we have two types of noisy modifications of the Bloch sphere: if $R > 1$ we interpret this as
depolarising noise just before the measurements at rate $p=(R-1)/R$, if $R < 1$ we interpret this as depolarising noise in the qubit preparations
at rate $1-R$. It will be convenient for us to use the following notation:

\medskip

{\bf Definition: Rescaling map.} Let $T_R$ denote the linear map corresponding to rescaling parameter $R$. Provided that
$R>0$ this map is invertible, and we denote the inverse by $T^{-1}_R$

\medskip

With these modifications of the single particle state space, our goal is to calculate the amount of noise
required before an entangling gate such as the CSIGN $C$ becomes non-entangling in the modified picture. The PPT
criterion can be applied to answer this question, because a two-particle operator $AB$ will be separable w.r.t. to
an $R$ rescaled single particle space iff $T^{-1}_R \otimes T^{-1}_R (AB)$ is both positive and PPT. Hence we need to work out the minimal
noise $N$ required before all the expression
\begin{equation}
T^{-1}_R \otimes T^{-1}_R \circ N \circ C \circ T_R \otimes T_R (\phi \otimes \psi) \label{rescale}
\end{equation}
is both positive and PPT for all possible input pure quantum product states $\phi \otimes \psi$.

Appendix III gives some
symmetry arguments that simplify this problem for local and joint depolarising models, and for local dephasing models.
In all these cases numerical search enabled us the minimal noise to make the outputs positive and PPT.
We expect from the numerics that the choices for $\phi$ and $\psi$ that need the most
noise to give output PPT and positive states can be chosen to be either $1/2(I+X)$ or $1/2(I+Z)$, however we have not yet been
able to confirm this analytically.
\begin{itemize}
\item {\bf Joint Depolarising.} We numerically obtained the graphs Fig(\ref{jointR}) and Fig(\ref{jointRasNoise}) for the tradeoff between rescaling and noise on the CSIGN required to ensure rescaled-separability. The graphs show that at a rescaling of $R=1.73$, corresponding to a local depolarisation rate of approx $42.2\%$ before measurements, joint depolarisation rates on the CSIGN of $53.6\%$, are sufficient to ensure rescaled separability. Hence despite the fact that the noisy CSIGN gates are slightly entangling, the noise on the measurements is sufficient to eradicate any effective entanglement.
\begin{figure}[t]
        \resizebox{6cm}{!}{\includegraphics{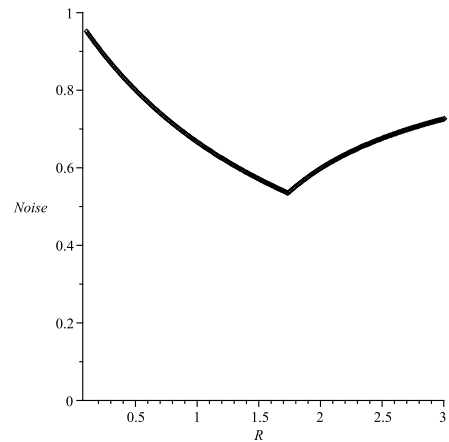}}
        \caption{Joint depolarisation noise required to remove CSIGN entanging power w.r.t. rescaled Bloch spheres. $R=1$ corresponds to no rescaling. We see that adding noise to
        the measurements in the device ($R>1$) allows the CSIGN gate to become separable w.r.t. the rescaled state space with slightly less noise.}
        \label{jointR}
\end{figure}
\begin{figure}[t]
        \resizebox{6cm}{!}{\includegraphics{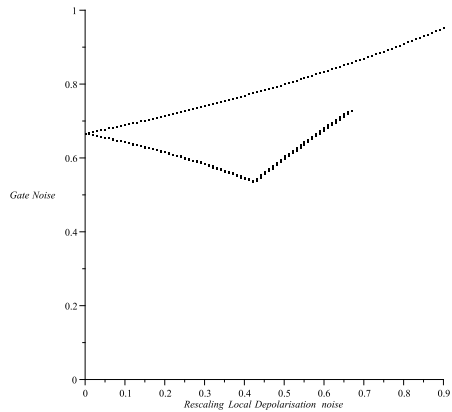}}
        \caption{Joint depolarisation with rescaled Bloch sphere. Here Fig(\ref{jointR}) is replotted with $R$ converted
        to the equivalent depolarisation rate on either preparation ($R <1$) or measurement ($R>1$). The curve that starts lower
        corresponds to $R>1$, the one starting higher to $R<1$. The vertical axis labels the noise required on the CSIGN gate
        to remove its entangling power w.r.t. to the rescaled Bloch sphere.}
        \label{jointRasNoise}
\end{figure}
\item {\bf Local depolarising.} We numerically obtained the graphs Fig(\ref{localR}) and Fig(\ref{localRasNoise}) for the tradeoff between rescaling and noise on the CSIGN required to ensure rescaled-separability. The graphs show that for rescaling of $R=1.16$, corresponding to a local depolarisation rate of approx $13.8\%$ before measurements, a local depolarisation rate of $39.5\%$, a is sufficient to ensure rescaled separability. Hence despite the fact that the noisy CSIGN gates are slightly entangling (one can see from the figures that the gate noise is lower than the value at $R=1$), the noise on the measurements is sufficient to eradicate any effective entanglement.
    \begin{figure}[t]
        \resizebox{6cm}{!}{\includegraphics{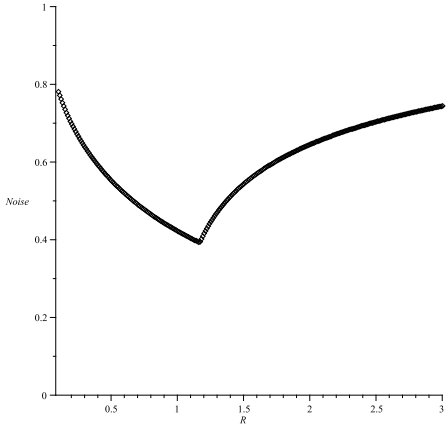}}
        \caption{Local depolarisation noise required to remove CSIGN entanging power w.r.t. rescaled Bloch spheres. $R=1$ corresponds to no rescaling. We see that adding noise to the measurements in the device ($R>1$) allows the CSIGN gate to become separable w.r.t. the rescaled state space with slightly less noise.}
        \label{localR}
\end{figure}
\begin{figure}[t]
        \resizebox{6cm}{!}{\includegraphics{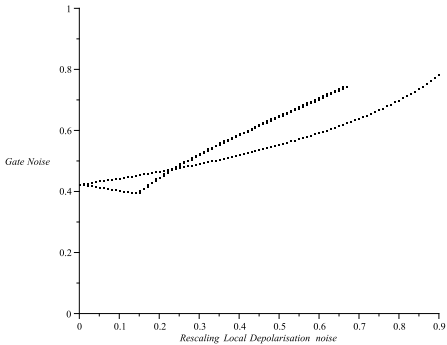}}
        \caption{Local depolarisation with rescaled Bloch sphere. Here Fig(\ref{localR}) is replotted with $R$ converted
        to the equivalent depolarisation rate on either preparation ($R <1$) or measurement ($R>1$). The curve that starts lower
        corresponds to $R>1$, the one starting higher to $R<1$. The vertical axis labels the noise required on the CSIGN gate
        to remove its entangling power w.r.t. to the rescaled Bloch sphere.}
        \label{localRasNoise}
\end{figure}
\item {\bf Local dephasing.} Interestingly it can be shown that for $R \neq 1$ only complete local dephasing (i.e. applying local
 Zs with independent $50\%$ probability on each side) is sufficient to
make the CSIGN separable w.r.t. rescaled cubes. If we choose $\phi=1/2(I+X)$ and $\psi=1/2(I+Z)$ we find that the output
of equation (\ref{rescale}) is proportional to (we drop normalisation):
\begin{equation}
     I \otimes I +(1-2p)R X \otimes I + I \otimes Z + {(1-2p) \over R} X \otimes Z \nonumber
\end{equation}
where $p$ is the probability of locally applying a $Z$ operator.
All terms in this operator commute, so it is easy to diagonalise. Two of the eigenvalues are:
\begin{eqnarray}
     (1-2p)(R-{1 \over R}) \nonumber \\
     (1-2p)({1 \over R}-R) \label{probs}
\end{eqnarray}
which cannot both be positive if $p \neq 1/2$ unless $R=1$.
This is very much analogous to our later findings for local dephasing on rescaled cubes. In fact,
the positivity of equations (\ref{probs}) is necessary just for positivity for the output.
Hence for $R\neq 1$ we cannot even have a valid probabilistic theory unless the local dephasing on
the CSIGN is total.
\end{itemize}

\section{Rescaled Cubes.}

The tradeoff investigated in the previous section can also be explored in the situation where
measurements are both restricted in direction and noisy. In magic-state architectures it could for example
be the case that a small amount of noise on the measurements or preparations could reduce the noise required to make the CSIGN gate separable w.r.t. the rescaled cubes. The noise models that we consider in this section are mixtures of Pauli operators, so by the arguments presented earlier it is sufficient to consider only one such input state.

There are two values of the rescaling parameter that have particular significance: values of
$R \geq 1/\sqrt{2}$ represent a rescaled cube that contain all the privileged pure `magic states' along the `T' and
`H' directions \cite{BK}, and values
of $R \geq 1/\sqrt{3}$ represent a rescaled cube that contain all the pure magic states in the `T' directions.
Hence we will pay special attention to $R$ values of $1/\sqrt{2}$ and $1/\sqrt{3}$. A rescaling value of $R=1/\sqrt{3}$ also has
further significance: it represents the local depolarising rate required to disentangle (in the usual quantum setting)
a Bell state. Hence as pure Bell states are cube-separable, this means that for $1 \geq R > 1/\sqrt{3}$ our device
can in principle also have access to entangled Bell diagonal states.

\subsection{Local Depolarisation for rescaled cubes}

\begin{figure}[t]
        \resizebox{6cm}{!}{\includegraphics{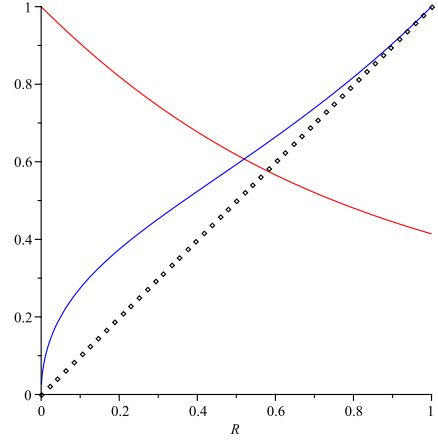}}
        \caption{The horizontal axis corresponds to $R$, the vertical to $r$. The decreasing curve
        corresponds to equation (\ref{xy}), the increasing curve to (\ref{xz}). The valid region
        must lie below both these curves - above them gives negative probabilities of outcomes. The dotted line is the line $R=r$.}
\label{curves}
\end{figure}

To analyse local depolarisation of rescaled cubes consider the sequence of transformations: (A) a rescaling by $R > 0$ (corresponding to noise on the measurements or preparations), (B) followed by a CSIGN, (C) followed by
a local noisy depolarisation operation (characterised by a rescaling $r < 1$), (D) followed by undoing
the rescaling $R$. Let us pick two input pure cube states
with Bloch vectors $(x,y,z)$ and $(A,B,C)$ to go into this sequence of transformations. This gives the following sequence of operators:
\begin{eqnarray} \nonumber \left(\begin{array}{cccc}
     1 & RA & RB & RC \\
     Rx & R^2xA & R^2xB & R^2xC \\
     Ry & R^2yA & R^2yB & R^2yC \\
     Rz & R^2zA & R^2zB & R^2zC
      \end{array}\right) \\ \nonumber \rightarrow
\left(\begin{array}{cccc}
     1 & R^2zA & R^2zB & RC \\
     R^2xC & R^2yB & -R^2yA & Rx \\
     R^2yC & -R^2xB & R^2xA & Ry \\
     Rz & RA & RB & R^2zC
      \end{array}\right) \\ \nonumber \rightarrow
\left(\begin{array}{cccc}
     1 & rR^2zA & rR^2zB & rRC \\
     rR^2xC & r^2R^2yB & -r^2R^2yA & r^2Rx \\
     rR^2yC & -r^2R^2xB & r^2R^2xA & r^2Ry \\
     rRz & r^2RA & r^2RB & r^2R^2zC
      \end{array}\right) \\ \nonumber \rightarrow
\left(\begin{array}{cccc}
     1 & rRzA & rRzB & rC \\
     rRxC & r^2yB & -r^2yA & r^2x/R \\
     rRyC & -r^2xB & r^2xA & r^2y/R \\
     rz & r^2A/R & r^2B/R & r^2zC
      \end{array}\right)
\end{eqnarray}
Our goal is hence to work out for a fixed value of $R$ the maximum $r$
(i.e. the minimal local depolarisation) required before the last state
in the sequence is cube separable for a given choice of $x,y,z,A,B,C \in \pm 1$.
Let us pick $x,y,z,A,B,C =+1$. The final state in the sequence becomes
\begin{eqnarray}
\left(\begin{array}{cccc}
     1 & rR & rR & r \\
     rR & r^2 & -r^2 & r^2/R \\
     rR & -r^2 & r^2 & r^2/R \\
     r & r^2/R & r^2/R & r^2
      \end{array}\right)
      \label{blah}
\end{eqnarray}
Again we need to make sure that all probabilities are positive. We numerically identified
 the following two probabilities as being important. Consider
the probability of getting the down outcome on both sides upon measuring $X \otimes Y$:
\begin{eqnarray}
{1 \over 4}(1-r^2-2rR) \geq 0 \nonumber \\
\Rightarrow \sqrt{1+R^2}-R \geq r \label{xy}
\end{eqnarray}
and the probability of getting the up on the first particle and down on the second particle upon measuring $X \otimes Z$:
\begin{eqnarray}
{1 \over 4}(1+rR-r-{r^2 \over R})\geq 0 \nonumber \\
\Rightarrow {R-1 + \sqrt{(R-1)^2+4/R} \over (2/R)} \geq r  \label{xz}
\end{eqnarray}
Both inequalities are plotted in figure (\ref{curves}). The first bound decreases monotonically as $R$ increases. Hence we know that choosing $R>1$ cannot lead to a  tightening of the bound that we have already obtained for $R=1$. Hence we may restrict our attention to rescaling factors corresponding to $R \leq 1$. The peak $r$ that satisfies both these inequalities corresponds to the intersection of the two curves, and is given by $1-r =39.2919\%$, at $1-R=47.9927\%$. However, the maximal $r$ local-depolarising level at which the CSIGN is still quantum-separable is $1-r=1 - {1 \over \sqrt{3}}=42.2649\%$. Hence even if $1-r =39.2919\%$ could correspond to a valid LHV model (which we do not believe from numerical investigations), the reduction in the fault tolerance upper bound would be marginal and comes at the cost of a fair amount of noise on the qubit preparations.

We in fact believe from numerical investigation that equation (\ref{xy}) corresponds to a valid LHV model only for $R \geq \approx 0.5449335$, corresponding to an $r$ value of $1-r=40.6095268\%$. However the $R$ values of $1/\sqrt{2},1/\sqrt{3}$ are covered by this region. At $R=\sqrt{1/2} \approx 0.7071$ we require a local depolarisation rate of $1-r = 1-(\sqrt{3}-1)/\sqrt{2} \approx 48.24\%$, and at $R=\sqrt{1/3} \approx 0.5773$ we require a local depolarisation rate of $1-r = 1-1/\sqrt{3} \approx 42.27\%$.
This last figure means that a stabilizer based computation with access to pure magic states in the T-direction becomes classically simulatable if the CSIGN gates and Pauli qubit preparations are subject to $42.27\%$ local depolarisation. This figure is effectively worse than bounds obtainable
from ordinary quantum separability: at local depolarisation rates of $1-r = 1-1/\sqrt{3}$ the CSIGN already loses the ability to
quantum entangle input qubit states.

\subsection{Joint Depolarisation for rescaled cubes}

\begin{figure}[t]
        \resizebox{6cm}{!}{\includegraphics{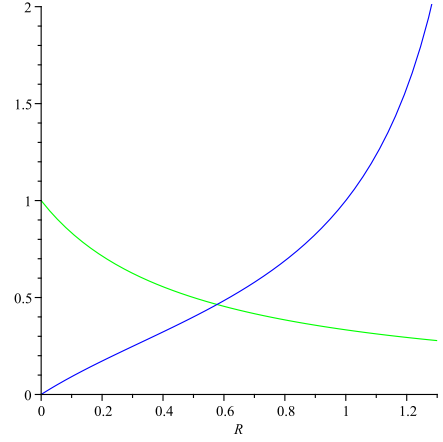}}
        \caption{The horizontal axis corrpesponds to $R$, the vertical to $1-\lambda=r$,
        where $\lambda$ is the joint depolarisation rate. The decreasing curve
        corresponds to equation (\ref{tdb1}), the increasing curve to (\ref{tdb2}). The valid region
        must be below both these curves - above them gives negative probabilities of outcomes. The intersection point of the two curves is not achievable by a LHV model according to numerical investigations. The intersection point is given by $(R,r)=(1/\sqrt{3},\sqrt{3}/(2+\sqrt{3}))\approx (0.58,0.46)$}.
\label{totaldepbox}
\end{figure}

For a total joint depolarising noise model the equivalent of equation (\ref{blah}) is:
\begin{eqnarray}
\left(\begin{array}{cccc}
     1 & rR & rR & r \\
     rR & r & -r & r/R \\
     rR & -r & r & r/R \\
     r & r/R & r/R & r
      \end{array}\right)
\end{eqnarray}
The probability of measuring down,down for a measurement of
$X \otimes Y$ is given by:
\begin{eqnarray}
{1 \over 4}(1-r-2rR) \geq 0 \nonumber \\
\Rightarrow r \leq {1 \over 2R+1} \label{tdb1}
\end{eqnarray}
Numerical investigations helped us to identify this as a key inequality. Another useful (but not as important) inequality
is the probability of getting up,down upon measurement of $Y \otimes Z$. This is given by:
\begin{eqnarray}
{1 \over 4}(1-r-(r/R)+rR) \geq 0 \nonumber \\
\Rightarrow r \leq {1 \over 1+(1/R)-R} \label{tdb2}
\end{eqnarray}
Both these inequalities are plotted in (\ref{totaldepbox}). Equation (\ref{tdb1}) shows that we may
only reach cube-separability with less noise if $R < 1$. Numerics show that the inequality (\ref{tdb1})
is achievable by a LHV model for values of $R$ down to $1-R = 1-{1 \over \sqrt{2}} \approx 29.29\%$, i.e. a noise value of:
\begin{eqnarray}
r = {1 \over \sqrt{2} + 1} \nonumber \\
\Rightarrow \lambda = 1 - {1 \over \sqrt{2} + 1} \approx 58.58 \%
\end{eqnarray}
This value of $R$ means that all pure magic states are admitted. Hence for a stabilizer based computation with access
to pure magic states, a local depolarisation rate on Pauli preparations of $29.29\%$ and a joint depolarisation rate
of $58.58\%$ on the CSIGN gates are sufficient to make the device efficiently classically simulatable. We can also
allow the machine access to slightly entangled noisy Bell states: as Bell states are separable w.r.t. unscaled cubes,
Bell states locally depolarised at a rate of $1-R$ will be separable w.r.t. the re-scaled cubes. However, local depolarisation
rates of $1-1/\sqrt{2}$ are not sufficient ($1-1/\sqrt{3}$ is required) to make a Bell state quantum separable.

\subsection{Local dephasing for rescaled cubes.}

It turns out that for any $R \neq 1$ the noisy CSIGN fails to be separable w.r.t. the rescaled cubes. The equivalent
of equation (\ref{blah}) is:
\begin{eqnarray}
\left(\begin{array}{cccc}
     1 & (1-2p)R & (1-2p)R & 1 \\
     (1-2p)R & (1-2p)^2 & -(1-2p)^2 & (1-2p)/R \\
     (1-2p)R & -(1-2p)^2 & (1-2p)^2 & (1-2p)/R \\
     1 & (1-2p)/R & (1-2p)/R & 1
      \end{array}\right) \nonumber
\end{eqnarray}
There is no LHV model that can match such a state - any product decomposition must have only
1s in the corner elements, but then the (1,2) and (4,2) elements must be identical, which they
are not if $R \neq 1$, unless the dephasing is total (i.e. $p=1/2$). Another way of seeing this
is that for measurements of $Z \otimes X$ giving down,up or down,down the probabilities are given by:
\begin{eqnarray}
     {1 \over 4}(1-1+(1-2p)R-(1-2p)/R)\geq 0 \\
     {1 \over 4}(1-1-(1-2p)R+(1-2p)/R)\geq 0
\end{eqnarray}
These inequalities cannot be satisfied if $R \neq 1$ unless the dephasing is total (i.e. $p=1/2$).
As these are conditions for a probability distribution, this means that no valid generalised theory
can be constructed on $R \neq 1$ re-scaled Bloch-cubes for a CSIGN with any partial amount of local dephasing.

\section{Conclusions and Discussion.}

We have explored the consequences of redefining the single particle state space as the set
of normalised positive operators with respect to restricted measurements.
This modification allows the consideration of non-quantum states and a different notion of entanglement, which can then
be fed into classical simulation methods for noisy quantum computers that rely on limited entanglement, such as \cite{HN}.

Measurements restricted to Pauli axes give a `cube' state space. We find that it is possible to write down noisy quantum gates that are (a) quantum-entangling and (b) non-Clifford operations, but are separable with respect to the modified cube state space. Hence the
method appears to provide efficiently classically simulatable regimes that are not covered straightforwardly by
existing methods (all qubit states, including non-Clifford ones, are contained within the cube). The operations represented
by the Choi-Jamiolkowski states in equation (\ref{counter})
are also not entanglement breaking, and so they also can certainly generate multiparticle entanglement. However, it may be the case
that repeated application of such operations cannot generate arbitrary long-range quantum entanglement. The precise differences in multi-particle
quantum-entangling abilities of cube-separable operations are something that warrant further investigation. Bell states are cube-separable and so in principle they can be included as a resource in the device, and this may help give the device an ability to generate multiparty entanglement.

The approach has strong links to studies of generalised theories and the issue of non-locality and non-contextuality versus efficient classical simulation. Indeed most of the technical calculations presented in this paper can be seen as attempts to construct a non-quantum theory where the dynamics are taken from noisy quantum theory but the single particle states spaces are different to quantum. For some noise models we find that we cannot modify the quantum state space for anything less than maximal noise on the quantum dynamics (e.g. for the locally dephased CSIGN gate).

One motivation for this work was to understand whether this approach can be useful for deriving upper bounds to
noisy non-classical computation thresholds. In relation to cube-separability, examples such as the operations represented by equation (\ref{counter}), and the fact that pure Bell states are cube-separable, suggest that for some noise models it should be possible to obtain tighter bounds.
However, for machines restricted to Pauli measurements, and for the natural noise models that we have considered, there is little or
no demonstrable reduction in the threshold upper bounds that can be obtained via the HN method or application of the Gottesman-Knill
theorem. Nevertheless, the approach may be of conceptual interest as it provides an alternative view of the non-classicality of magic-state quantum computers - instead of magic-state or conventional quantum entanglement being viewed as the non-classical resource, the ability to generate non-local correlations or non-positive dynamics of the cubes is seen as essential to the non-classicality.

In the case where measurements are not restricted to a particular direction but they are slightly noisy, we have obtained curves
showing the tradeoff between noise on measurements and the gate noise required to make the CSIGN separable with respect to
the rescaled Bloch sphere. In these cases we do observe that adding small amounts
of depolarisation noise to the measurements does require less noise on the CSIGN gates to induce
separability, although the effect is slight (e.g. Fig.(\ref{jointRasNoise}, Fig.(\ref{localRasNoise}))). However it is
interesting to note that adding noise to qubit preparations does not have the same effect, in contrast it requires {\it more} noise on the CSIGN gates. It is also interesting to note that for rescaling factors $R\neq1$ a locally dephased CSIGN gate cannot give dynamics in a valid generalised probabilistic theory unless the dephasing is maximal.

Many of the questions raised can be naturally generalised to higher dimensions: our restricted measurement sets have been
constructed by selecting a single measurement and generating a full set using a group of single particle unitaries, and this process
can extend naturally to higher dimensions. In particular when the sets of measurements are both tomographically complete and `operator compatible' it will also
hold that non-classical computation requires the ability to generate non-local correlations. This applies to all fault tolerance schemes that use only generalised Pauli measurements in higher dimensions.

\section{Acknowledgments}

This work is supported by EU STREP `Corner', a University of Strathclyde Starter grant, and an EPSRC PhD studentship.
We thank Jon Barrett, David Gross,  Miguel Navascues and Rob Spekkens for discussions, Earl Campbell for constructive criticism.

\begin{widetext}

\section{APPENDIX: Relevant local hidden variable models.}

In this section we mostly ignore normalisation factors, and represent product cube states by notation such as:
\begin{eqnarray}
\left(\begin{array}{cccc}
     1 & 1 & 1 & 1 \\
     1 & . & . & . \\
     1 & . & . & . \\
     1 & . & . & .
      \end{array}\right)
\end{eqnarray}
where the dots ``.'' indicate that the element is the product of the element at the beginning of the row and
top of the column. We construct a number of LHVs in sequence. Later LHVs often are built using earlier ones.
Most of these models were developed by trial and error in order to match the inequalities needed for the output of the noisy CSIGN gate
to give valid probabilities for Pauli measurements.
\begin{enumerate}
 \item \begin{eqnarray} \nonumber \left(\begin{array}{cccc}
     1 & 0 & 0 & 0 \\
     0 & 1 & 1 & 1 \\
     0 & 1 & 1 & 1 \\
     0 & 1 & 1 & 1
      \end{array}\right) = {1 \over 2} \left(\begin{array}{cccc}
     1 & 1 & 1 & 1 \\
     1 & . & . & . \\
     1 & . & . & . \\
     1 & . & . & .
      \end{array}\right) +
      {1 \over 2} \left(\begin{array}{cccc}
     1 & -1 & -1 & -1 \\
     -1 & . & . & . \\
     -1 & . & . & . \\
     -1 & . & . & .
      \end{array}\right)
\end{eqnarray}

\item \begin{eqnarray} \nonumber \left(\begin{array}{cccc}
     1 & 0 & 0 & 0 \\
     0 & 1 & -1 & 0 \\
     0 & -1 & 1 & 0 \\
     0 & 0 & 0 & 0
      \end{array}\right) \propto \sum_{q,r,s,t \in \pm 1} \left(\begin{array}{cccc}
     1 & 1 & -1 & q \\
     1 & . & . & . \\
     -1 & . & . & . \\
     r & . & . & .
      \end{array}\right) +
      \left(\begin{array}{cccc}
     1 & -1 & 1 & s \\
     -1 & . & . & . \\
     1 & . & . & . \\
     t & . & . & .
      \end{array}\right)
\end{eqnarray}

\item \begin{eqnarray} \nonumber \left(\begin{array}{cccc}
     1 & 0 & 0 & 0 \\
     0 & 0 & -1 & 0 \\
     0 & -1 & 0 & 0 \\
     0 & 0 & 0 & 0
      \end{array}\right) \propto \sum_{p,q,r,s \in \pm 1} \left(\begin{array}{cccc}
     1 & p & q & r \\
     -q & . & . & . \\
     -p & . & . & . \\
    s & . & . & .
      \end{array}\right)
\end{eqnarray}

\item Joint depolarising noise LHV:
\begin{eqnarray} \nonumber \left(\begin{array}{cccc}
     1 & 1/3 & 1/3 & 1/3 \\
     1/3 & 1/3 & -1/3 & 1/3 \\
     1/3 & -1/3 & 1/3 & 1/3 \\
     1/3 & 1/3 & 1/3 & 1/3
      \end{array}\right) = {1 \over 3} \left(\begin{array}{cccc}
     1 & 1 & 1 &  1 \\
     1 & . & . & . \\
     1 & . & . & . \\
     1 & . & . & .
      \end{array}\right) +
      {2 \over 3} \left(\begin{array}{cccc}
     1 & 0 & 0 & 0 \\
     0 & 0 & -1 & 0 \\
     0 & -1 & 0 & 0 \\
     0 & 0 & 0 & 0
      \end{array}\right)
\end{eqnarray}

\item LHVs required for later local dephasing model:
\begin{eqnarray} \nonumber \left(\begin{array}{cccc}
     1 & 1 & 1 & 1 \\
     0 & 0 & 0 & 0 \\
     0 & 0 & 0 & 0 \\
     1 & 1 & 1 & 1
      \end{array}\right) = {1 \over 2} \left(\begin{array}{cccc}
     1 & 1 & 1 &  1 \\
     1 & . & . & . \\
     1 & . & . & . \\
     1 & . & . & .
      \end{array}\right) +
      {1 \over 2} \left(\begin{array}{cccc}
     1 & 1 & 1 &  1 \\
     -1 & . & . & . \\
     -1 & . & . & . \\
     1 & . & . & .
      \end{array}\right)
\end{eqnarray}

\begin{eqnarray} \nonumber \left(\begin{array}{cccc}
     1 & 0 & 0 & 1 \\
     1 & 0 & 0 & 1 \\
     1 & 0 & 0 & 1 \\
     1 & 0 & 0 & 1
      \end{array}\right) = {1 \over 2} \left(\begin{array}{cccc}
     1 & 1 & 1 &  1 \\
     1 & . & . & . \\
     1 & . & . & . \\
     1 & . & . & .
      \end{array}\right) +
      {1 \over 2} \left(\begin{array}{cccc}
     1 & -1 & -1 &  1 \\
     1 & . & . & . \\
     1 & . & . & . \\
     1 & . & . & .
      \end{array}\right)
\end{eqnarray}

\item LHV for local dephasing. The following is a valid LHV provided that $1-2(1-2p)-(1-2p)^2 \geq 0$ is satisfied:
\begin{eqnarray} \nonumber \left(\begin{array}{cccc}
     1 & 1-2p & 1-2p & 1 \\
     1-2p & (1-2p)^2 & (1-2p)^2 & 1-2p \\
     1-2p & (1-2p)^2 & (1-2p)^2 & 1-2p \\
     1 & 1-2p & 1-2p & 1
      \end{array}\right) = (1-2(1-2p)-(1-2p)^2) \left(\begin{array}{cccc}
     1 & 0 & 0 & 1 \\
     0 & 0 & 0 & 0 \\
     0 & 0 & 0 & 0 \\
     1 & 0 & 0 & 1
      \end{array}\right) + \nonumber \\
      (1-2p)\left\{ \left(\begin{array}{cccc}
     1 & 0 & 0 & 1 \\
     1 & 0 & 0 & 1 \\
     1 & 0 & 0 & 1 \\
     1 & 0 & 0 & 1
      \end{array}\right) +
       \left(\begin{array}{cccc}
     1 & 1 & 1 & 1 \\
     0 & 0 & 0 & 0 \\
     0 & 0 & 0 & 0 \\
     1 & 1 & 1 & 1
      \end{array}\right) \right\} +
      (1-2p)^2 \left(\begin{array}{cccc}
     1 & 0 & 0 & 0 \\
     0 & 1 & -1 & 0 \\
     0 & -1 & 1 & 0 \\
     0 & 0 & 0 & 0
      \end{array}\right) \nonumber
\end{eqnarray}

\item LHV for local depolarising. The following is a valid LHV provided that $1-2(1-p)-(1-p)^2 \geq 0$ is satisfied:
\begin{eqnarray} \nonumber \left(\begin{array}{cccc}
     1 & 1-p & 1-p & 1-p \\
     1-p & (1-p)^2 & (1-p)^2 & (1-p)^2 \\
     1-p & (1-p)^2 & (1-p)^2 & (1-p)^2 \\
     1-p & (1-p)^2 & (1-p)^2 & (1-p)^2
      \end{array}\right) = (1-2(1-p)-(1-p)^2) \left(\begin{array}{cccc}
     1 & 0 & 0 & 0 \\
     0 & 0 & 0 & 0 \\
     0 & 0 & 0 & 0 \\
     0 & 0 & 0 & 0
      \end{array}\right) + \nonumber \\
      ((1-p)-(1-p)^2)\left\{ \left(\begin{array}{cccc}
     1 & 1 & 1 & 1 \\
     0 & 0 & 0 & 0 \\
     0 & 0 & 0 & 0 \\
     0 & 0 & 0 & 0
      \end{array}\right) +
       \left(\begin{array}{cccc}
     1 & 0 & 0 & 0 \\
     1 & 0 & 0 & 0 \\
     1 & 0 & 0 & 0 \\
     1 & 0 & 0 & 0
      \end{array}\right) \right\} +
      3(1-p)^2 \left(\begin{array}{cccc}
     1 & 1/3 & 1/3 & 1/3 \\
     1/3 & 1/3 & -1/3 & 1/3 \\
     1/3 & -1/3 & 1/3 & 1/3 \\
     1/3 & 1/3 & 1/3 & 1/3
      \end{array}\right) \nonumber
\end{eqnarray}

\end{enumerate}

\section{Appendix II: Noiseless CSIGN is not a cube-positive operation for state spaces bigger than
the Bloch sphere.}

Let us pick input pure cube states $x=1,C=-1,z=A=1,y=B=1$ for a noiseless CSIGN gate.
Applying this choice to equation (\ref{cond}) gives -1/2, which is not a valid probability. So the theory is
not self consistent if the CSIGN is admitted as a 2-cube operation acting on all cube states. This tells us that we would need to add noise to the CSIGN not
only to make the output cube-separable, but even just to make the output correspond
to a valid probability distribution when measured using Pauli operators.

In fact the same argument can be made to go a little further. One might consider more general
situations in which other sets of measurements (in addition to the $X,Y,Z$ measurements) are used
to define the set of allowed operators - perhaps by picking an additional projective measurement
in some direction and considering what measurements would be obtained by allowing the Clifford group (or another group
containing it) to act upon it by conjugation. This group could then
be the allowed single qubit unitaries in a proposal for a quantum computer. The set of measurements obtained using
this group would then define a set of `states' that encloses
the Bloch sphere. However, as the following argument shows, if this set of states is
required to (a) contain the Bloch sphere within it, and (b) Pauli measurements are allowed, then the only consistent
set allowing the CSIGN gate is the Bloch sphere itself. Equation (\ref{cond}) gives the requirement:
\begin{equation}
{1 \over 4} ( 1 + xC - yB - zA) \geq 0
\end{equation}
as this needs to be a valid probability. But let us fix $V=(C,-B,-A)$ to be any unit vector from the Bloch sphere (as our set of
allowed states contains the Bloch sphere, we can do this, as $A,B,C$ can be the components
of any unit vector). Denoting
$v = (x,y,z)$, we have:
\begin{equation}
 1 + v.V \geq 0
\end{equation}
But the only way that this can be true for all possible choices of $V$ is if $v$ is a vector of less than unit norm, i.e.
the Bloch sphere. Hence we see that the CSIGN gate cannot output only positive states for any set of single particle states
that is strictly bigger than the Bloch sphere, and hence cannot even be separable.

\section{Appendix III: Rescaled Bloch sphere symmetry simplifications.}

Consider the following sequence of transformations: (A) pick two pure quantum states from the Bloch sphere $\psi$ and $\phi$,
(B) rescale by a factor $R>0$ in accordance with the noise on the measurements or preparations, (C) apply a
CSIGN gate $C$, (C) apply the noise model $N$ under consideration, (D) apply the inverse of the rescaling.
The total sequence is: $(T^{-1}_R \otimes T^{-1}_R) \circ N \circ C \circ (T_R \otimes T_R) (\psi \otimes \phi)$.
Local unitaries after this transformation will not affect the positivity or PPT characteristics of the output.

Let us consider applying local rotations about the $Z$ axis on each particle.
All noise models that we will consider commute (as CP maps) with the application of rotations about the $Z$ axis
- we will only consider local depolarising, dephasing, or joint depolarising.
As local rotations about the $Z$ axis also commute with $T^{-1}_R$, $T_R$, and the CSIGN gate, this means the rotations
can be pushed all the way through onto the input states. Hence we are free
to pick $\psi$ and $\phi$ up to arbitrary rotations about the $Z$ axis. We will hence use this freedom to choose input pure states with
no $Y$ component.

Under the noise models each Pauli product term in the expansion will acquire a factor
$0 \leq r \leq 1$, which can in principle vary for different Pauli terms. However, for the purposes
of the following argument we may simply set it to be constant, as it would be for the joint depolarisation
noise model. Then the sequence of transformations will be represented as:
\begin{eqnarray} \nonumber \left(\begin{array}{cccc}
     1 & R\cos(\phi) & 0 & R\sin(\phi) \\
     R\cos(\theta) &  R^2 \cos(\theta) \cos(\phi) & 0 &  R^2\cos(\theta)\sin(\phi) \\
     0 & 0 & 0 & 0 \\
     R\sin(\theta) & R^2\sin(\theta)\cos(\phi) & 0 & R^2\sin(\theta)\sin(\phi)
      \end{array}\right) \\ \nonumber \rightarrow
\left(\begin{array}{cccc}
     1 & R^2\sin(\theta)\cos(\phi) & 0 & R\sin(\phi) \\
     R^2\cos(\theta)\sin(\phi) &  0 & 0 & R\cos(\theta)  \\
     0 & 0 & R^2 \cos(\theta) \cos(\phi) & 0 \\
     R\sin(\theta) & R\cos(\phi) & 0 & R^2\sin(\theta)\sin(\phi)
      \end{array}\right) \\ \nonumber \rightarrow
\left(\begin{array}{cccc}
     1 & rR^2\sin(\theta)\cos(\phi) & 0 & rR\sin(\phi) \\
     rR^2\cos(\theta)\sin(\phi) &  0 & 0 & rR\cos(\theta)  \\
     0 & 0 & rR^2 \cos(\theta) \cos(\phi) & 0 \\
     rR\sin(\theta) & rR\cos(\phi) & 0 & rR^2\sin(\theta)\sin(\phi)
      \end{array}\right) \\ \nonumber \rightarrow
\left(\begin{array}{cccc}
     1 & rR\sin(\theta)\cos(\phi) & 0 & r\sin(\phi) \\
     rR\cos(\theta)\sin(\phi) &  0 & 0 & {r \over R}\cos(\theta)  \\
     0 & 0 & r \cos(\theta) \cos(\phi) & 0 \\
     r\sin(\theta) & {r \over R}\cos(\phi) & 0 & r\sin(\theta)\sin(\phi)
      \end{array}\right)
\end{eqnarray}
Partial transposition on the second system applies a minus sign to the third column,
and so our goal is to show that for both choices of sign the operators:
\begin{eqnarray}
\left(\begin{array}{cccc}
     1 & rR\sin(\theta)\cos(\phi) & 0 & r\sin(\phi) \\
     rR\cos(\theta)\sin(\phi) &  0 & 0 & {r \over R}\cos(\theta)  \\
     0 & 0 & \pm r \cos(\theta) \cos(\phi) & 0 \\
     r\sin(\theta) & {r \over R}\cos(\phi) & 0 & r\sin(\theta)\sin(\phi)
      \end{array}\right) \nonumber
\end{eqnarray}
are positive. A few observations concerning symmetries are helpful in this regard:
(a) adding $\pi$ to $\theta$ changes the sign of the $\sin(\theta)$ and $\cos(\theta)$. But it can be verified that this can
also be achieved by applying a final local unitary
$Y \otimes Z$. Similarly $\phi$ changing by $\pi$ is equivalent to applying $Z \otimes Y$. As unitaries
don't change the spectra, we may restrict our attention to $\theta,\phi \in [0,\pi]$,
(b) similarly it can be shown that applying $Z \otimes I$ is equivalent to changing $\Delta:=\theta - \pi/2$ to $-\Delta$,
and $I \otimes Z$ is equivalent to changing $\Delta:=\phi - \pi/2$ to $-\Delta$, and hence
w.l.o.g. we may restrict attention to $\theta,\phi \in [0,\pi/2]$.

\end{widetext}


\begin{thebibliography}{99}
\bibitem{foot} In any magic state circuit diagram each magic state will be next to or close to a CSIGN gate (or equivalent), and so the noise that we place on the CSIGN gates can be shifted onto the magic-states. The magic states will typically become Clifford states (convex mixtures of Stabilizer pure states) with much less noise than required to make the CSIGN quantum- or cube- separable. This depends strongly upon the noise model and precise magic state however - for dephasing noise models for example, there are qubit states (close to $Z$ eigenstates) that require maximal dephasing to become Clifford states.
\end{thebibliography}
\end{document}